\documentclass[11pt,a4paper,pre, showpacs]{revtex4}
\usepackage{graphicx}
\usepackage{amssymb}
\usepackage{epstopdf}
\usepackage{amsmath}
\usepackage{amsfonts}
\begin{document}

\title{Uncovering the topology of configuration space networks}
\author{David Gfeller$^1$}
\author{Paolo De Los Rios$^1$}
\author{David Morton de Lachapelle$^1$}
\author{Guido Caldarelli$^{2,3}$}
\author{Francesco Rao$^2$}
\affiliation{$^1$Laboratoire de
Biophysique Statistique, SB/ITP, Ecole Polytechnique F\'ed\'erale de Lausanne (EPFL),
CH-1015, Lausanne, Switzerland\\
$^2$Museo Storico della Fisica e Centro Studi e Ricerche "E. Fermi", 00184 Rome, Italy\\
$^3$SMC, INFM-CNR, Dipartimento di Fisica,
Universit\`a 'Sapienza', Piazzale Aldo Moro 5, 00185 Rome, Italy}

\date{\today}

\begin{abstract}
The configuration space network (CSN) of a dynamical system is an 
effective approach to represent the ensemble of configurations sampled 
during a simulation and their dynamic connectivity. To elucidate the 
connection between the CSN topology and the underlying free-energy 
landscape governing the system dynamics and thermodynamics, an analytical solution is provided to explain the heavy tail of the degree distribution, neighbor connectivity and clustering coefficient. 
This derivation allows to understand the universal CSN network 
topology observed in systems ranging from a simple quadratic well to the native state of the beta3s peptide and a 2D lattice heteropolymer.
Moreover CSN are shown to fall in the general class of complex networks described by the fitness model.

\end{abstract}


\maketitle
\clearpage

\section{Introduction}

 The use of complex networks and graph theory to describe complex 
systems ranging from the WWW to protein interaction networks is by now 
well-established (different book and reviews are available about 
this topic \cite{Albert2002,Dorogovstev2002,Newman2003-2,Boccaletti2006,Caldarelli2007}). A large class of these systems attains complexity 
by means of their internal dynamics \cite{cgc2007}, which is often 
revealed by computer 
simulations. One example of such systems is the folding of proteins for 
which simulations have been extensively used in structural biology. 
Nowadays several Molecular Dynamics (MD) softwares (CHARMM~\cite{Brooks1983}, 
GROMACS~\cite{Berendsen1995}, AMBER~\cite{Pearlman1995}) are available 
to probe in real time the dynamics of folding/unfolding. 
However, because of the large number of degrees of freedom \cite{Du1998} 
involved in the process, the results of MD simulations form themselves a 
highly complex system. As a consequence a detailed, unbiased description 
of the free-energy landscape underlying the thermodynamics and kinetics 
cannot not easily be extracted.

To tackle this complexity, new
approaches based on complex networks have been recently introduced 
showing that a network description is effective for the analysis and 
visualization of simulation results. In \cite{Scala2001} for instance, 
the topology of the configurations of a short lattice
polymer has been mapped onto a network.
Doye and Massen have also applied graph analysis to study the organization
of the potential energy minima in a Lennard-Jones cluster of atoms \cite{Doye2002, Doye2005}.In another work the concept of disconnectivity graphs has been used to analyze the free-energy of a tetrapeptide and a $\beta$-hairpin \cite{Krivov2002, Krivov2004}.
Finally the free-energy 
landscape of a three-stranded $\beta$-sheet (beta3s) and alanine dipeptide 
sampled by MD simulations have been represented as a configuration 
space network (CSN) \cite{Rao2004, Gfeller2007-1}.

Given the time evolution of a 
dynamical system, the CSN represents the ensemble
of micro-states (configurations) sampled during a simulation, and their dynamic connectivity. In this representation, nodes are system configurations and links are direct transitions between the configurations sampled
during the simulation.
The CSN topology shares several features
with other networks representing systems as different as cell
function \cite{Barabasi2004}, scientific collaborations \cite{Newman2001-2} and
the WWW \cite{Gibson1998}.
In particular it has been shown \cite{Rao2004} that the degree 
distribution of the beta3s peptide CSN exhibits a heavy tail well 
approximated by a power-law and a disassortative behavior for the 
average neighbor connectivity distribution. Moreover the clustering 
coefficient presents a decay compatible with a $1/k$ function 
for large values of the degree, which has
been interpreted as the presence of a hierarchical organization of the nodes \cite{Ravasz2003}.
Recently, the connection between the CSN topological clusters and  free-energy
basins has been explored and an analytical solution for the node weight distribution observed in CSNs has been provided
with the help of simple energy landscape models \cite{Gfeller2007-1}. 
Following these lines of research, the challenge is now to find the connection
between network topology, system dynamics and free-energy landscape organization. 
This work focuses on the degree distribution $P(k)$, the average 
neighbor connectivity $K_{nn}(k)$ and the clustering 
coefficient $C(k)$ observed in CSNs. Several studies 
\cite{Barabasi1999,Watts1998,PastorSatorras2001,Newman2003-1} have shown 
that the analysis of the above three distributions is an important 
step towards the understanding of the network organization and 
architecture. The results presented below provide the first rationale 
for the origin of several unexplained properties of CSNs.

The paper is organized as follows. Section II describes in detail how CSNs are built. Section III shows how the degree distribution, the average neighbor degree and the clustering coefficient relate to the free-energy landscape. In section IV an analytical derivation and simulation results are presented for the quadratic well model. Then the CSNs obtained from beta3s peptide and lattice heteropolymer simulations are analyzed in Section V. Finally, the connection between CSNs and the fitness model is discussed in Section VI and conclusions are presented in Section VII.

\section{Configuration Space Networks}
The simulation of a dynamical system, like a peptide or a protein, results in a time series of snapshots representing the dynamics.
The CSN of this kind of processes gives a synthetic view of the configurations and transitions observed during the simulation. System configurations are the nodes and a link is placed between two nodes if they appear consecutively in the time series. The time step between two snapshots $t_{s}$ (usually called configuration saving time) is a free parameter: $t_{s}= i_{t} M$, where $i_{t}$ is the integration time step for the simulation and $M$ is the number of microscopic steps between two snapshots. When $M$ approaches 1, only configurations spatially close to each other are connected together. Therefore a link is a temporal relation between configurations and changing $M$ changes the set of links.

The weight of a link $w_{ij}$ represents the number of direct transitions from node $i$ to $j$. Similarly, the weight $w_i$ of a node is given by the number of times a configuration has been visited. The weight distribution of CSN has been discussed in a previous work \cite{Gfeller2007-1}.

The degree of a node is defined as the number of links including loops (edges 
to itself), corresponding to the number of configurations accessed in $M$ steps during the dynamics.
Because of finite-time simulations, the CSN is a directed network: if the system visited node $j$ $M$ steps after node $i$, the converse is not 
automatically true. Hence $k^{in}$ and $k^{out}$ are not always equal. 
However, the asymmetry of the links is weak for two reasons. Firstly, 
the simulation is run long enough to almost ensure $w_{i\rightarrow j}
=w_{j\rightarrow i}$ (which is in fact equivalent to detailed balance). 
Secondly, the total weight of the incoming links has to be equal to 
the total weight of the outgoing links by construction of the network.

In the following, the degree of a node, $k_i$, is defined as the out-degree $k^{out}_i$. Similarly the average neighbor degree $K_{nn}(k)$ is the average out-degree of the neighbors of the nodes with degree $k$. The out-degree correlation between connected nodes is further characterized by  
the assortativity coefficient $q$ \cite{Newman2002}. Finally the clustering coefficient is computed as the total number of 3-steps cycles (triangles) 
starting at node $i$ ($N^{\bigtriangleup}_i$), 
divided by the maximum number of 3-steps cycles one can have in the 
considered graph: 
$c_i=\frac{N^{\bigtriangleup}_i}{k^{out}_ik^{in}_i}$. 

\section{Analytical approach}
\label{ana}

As already mentioned the ultimate goal is to understand the relation between the network topology and the free-energy landscape. Unfortunately the degree of a node for instance cannot be computed from the knowledge of the energy landscape for any $M$. However, large values of $M$ correspond to a random sampling of the landscape (uncorrelated exploration). In this case an analytical approach can be carried out. The probability density on the free-energy landscape is given by $W ({\bf x}) = W_o\exp(-U({\bf x}))$. $U({\bf x})$ (in $k_{B}T$ units) is the multi-dimensional energy potential and $W_o$ the correct normalization. As shown further in this article, uncorrelated exploration of the free-energy landscape is relevant for several simulation procedures. In the space ${\bf x}$, where $D$ is the dimension of the landscape, the system configurations (i.e., CSN nodes) are defined as hyper-cubic cells of size $a^{D}$. The probability to visit a configuration at ${\bf x_{1}}$ at a given time is $P({\bf x_{1}})=a^DW_0\exp(-U({\bf x_1}))$ and the expected number of times two configurations at position ${\bf x_{1}}$ and $\bf x_{2}$ are visited consecutively is given by: 
\begin{equation}
W(\mathbf{x_1}, \mathbf{x_2})=W_{N}e ^{-U(\mathbf{x_2})-U(\mathbf{x_1})}
\label{wij}
\end{equation}
where $W_{N}=N a^{2D}W_{o}^{2}$, $N$ is the total number of snapshots and $a$ is chosen small enough, such that $\exp(-U(\mathbf{x}))$ is almost constant on each cell. The expression above predicts link weights. To compute the degree, the quantity of interest is the probability $P(\mathbf{x_1}, \mathbf{x_2})$ to have a link between two configurations (no matter how often the link has been visited). Assuming that the probability distribution of visiting $s$ times the node at $\bf x_{1}$ is peaked around its average value $P(\mathbf{x_1})$, $P(\mathbf{x_1}, \mathbf{x_2})$ is evaluated as one minus the probability to have no links:
\begin{equation}
P(\mathbf{x_1}, \mathbf{x_2})=1-\left(1-P(\mathbf{x_2})\right)^{NP(\mathbf{x_1})}\approx 1-e^{-NP(\mathbf{x_1})P(\mathbf{x_2})}
\label{start}
\end{equation}
where the second equality holds in the limit of small $P(\mathbf{x_2})$, which is true if the number of configurations is large. $NP(\mathbf{x_1})$ is the expected number of times the configuration at $\mathbf{x_1}$ has been visited and $1-P(\mathbf{x_2})$ is the probability not to visit the configuration at $\mathbf{x_2}$. Eq. \ref{start} is an approximation. An exact expression would require to sum the probability of visiting $s$ times node $\mathbf{x_1}$ multiplied by the probability of never visiting $\mathbf{x_2}$ right after $\mathbf{x_1}$, i.e $(1-P(\mathbf{x_2}))^s$, excluding the cases in which $\mathbf{x_1}$ has been visited several times consecutively. However it is very difficult to express this in a simple form, and approximations are needed to proceed with further analytical calculations. From Eq. \ref{start}, two asymptotic behaviors can be derived:

\begin{enumerate}
\item If $NP(\mathbf{x_2})P(\mathbf{x_1})$ is large, then $P(\mathbf{x_1}, \mathbf{x_2})\approx 1$. This is the \emph{saturation regime} since $\mathbf{x_1}$ and $\mathbf{x_2}$ are almost certainly connected
\item If $NP(\mathbf{x_2})P(\mathbf{x_1})$ is small, the \emph{sparse regime} is reached which describes low probability connections. In this regime a $n^{\text{th}}$ expansion is meaningful:
\begin{equation}
P^{(n)}(\mathbf{x_1}, \mathbf{x_2})=\sum_{j=1}^{n}\frac{1}{j!}\left[NP(\mathbf{x_1})P(\mathbf{x_2})\right]^j(-1)^{j+1}
\label{taylor}
\end{equation}

\end{enumerate}
The first term in Eq.~\ref{taylor} is equal to $W(\mathbf{x_1}, \mathbf{x_2})$ in Eq. \ref{wij}. Taking only the first term in the sum corresponds to the case in which links are distributed avoiding as much as possible the presence of double links. If $NP(\mathbf{x_2})P(\mathbf{x_1})\ll 1$ it provides a good approximation of the sum, while for $NP(\mathbf{x_2})P(\mathbf{x_1})\approx 1$ it slightly overestimates the real probability to have a link between two nodes.
Applying the considerations above, an approximation of $P(\mathbf{x_1}, \mathbf{x_2})$ is given by:
\begin{eqnarray}
E^{(n)}(\mathbf{x_1}, \mathbf{x_2})=\text{min}\left[1,P^{(n)}(\mathbf{x_1}, \mathbf{x_2})\right]
\label{unif}
\end{eqnarray}

Eq. \ref{unif} defines a probability to have a link between two nodes, depending only on a parameter associated with each node (in this case the energy $-U({\bf x})$). Such systems have been first described as the fitness model in \cite{Caldarelli2002} and generalized in \cite{Boguna2003, Servedio2004} (see Section VI).
The degree of a node at $\mathbf{x}$, its average neighbor degree and the expected number of triangles the node is part of are then given by the following three expressions:
\begin{eqnarray}
k(\mathbf{x})=\frac{1}{a^D}\int_V\text{d}\mathbf{x_1}E^{(n)}(\mathbf{x}, \mathbf{x_1})
\label{deg}
\end{eqnarray}
\begin{eqnarray}
K_{nn}(\mathbf{x})=\frac{1}{a^D} \frac{1}{k(\mathbf{x})}\int_V\text{d}\mathbf{x_1}E^{(n)}(\mathbf{x}, \mathbf{x_1})k(\mathbf{x_1})
\label{knn}
\end{eqnarray}
\begin{eqnarray}
N^{\triangle}(\mathbf{x})=\frac{1}{a^{2D}}\int_V\int_V\text{d}\mathbf{x_1}\text{d}\mathbf{x_2}E^{(n)}(\mathbf{x}, \mathbf{x_1})E^{(n)}(\mathbf{x}, \mathbf{x_2})E^{(n)}(\mathbf{x_1}, \mathbf{x_2})
\label{tr}
\end{eqnarray}

Finally, assuming continuous approximation, the degree distribution reads:

\begin{equation}
P(k)\sim \int_V \text{d}^D\mathbf{x} \delta(k-k(\mathbf{x}))
\label{distr}
\end{equation}

Inverting Eq. \ref{deg} and inserting it into Eq.~\ref{knn} and Eq.~\ref{tr} gives the average neighbor connectivity $K_{nn}(k)$ and the clustering coefficient $C(k)$, respectively:  

\begin{equation}
K_{nn}(k)= \frac{1}{a^D} \frac{1}{k}\int_V \text{d}\mathbf{x_1}E^{(n)}(\mathbf{x(k)}, \mathbf{x_1})k(\mathbf{x_1})
\label{knn.k}
\end{equation}
\begin{equation}
C(k)=\frac{1}{k^2a^{2D}}\int_V \int_V\text{d}\mathbf{x_1}\text{d}\mathbf{x_2}E^{(n)}(\mathbf{x(k)}, \mathbf{x_1})E^{(n)}(\mathbf{x(k)}, \mathbf{x_2})E^{(n)}(\mathbf{x_1}, \mathbf{x_2})
\label{c.k}
\end{equation}

\section{Quadratic well}

In general the free-energy landscapes of real systems are extremely complex, so that even writing down a mathematical expression is often impossible. However, close to the minimum of a basin (corresponding to configurations visited several times), such systems can often be approximately described by means of a Taylor expansion of the potential, whose first term is harmonic.

Therefore the quadratic well is a good benchmark to understand more complex CSNs \cite{Gfeller2007-1}, in particular for nodes near the minimum of an energy basin.
In 2 dimensions, the potential is given by:
\begin{equation}
U(x,y)=\frac{1}{2}(x^2+y^2)=\frac{1}{2}r^{2}
\label{uxy}
\end{equation}
Using radial coordinates and introducing Eq. \ref{uxy} in Eq. \ref{unif} with $n=1$  gives ($W_0=\frac{1}{2\pi}$):
\begin{eqnarray}
P^{(1)}(\mathbf{x_1}, \mathbf{x_2})=1 & \Leftrightarrow & \frac{(2\pi)^2}{a^4N}=\exp\left(-\frac{1}{2}(r_1^2+r_2^2)\right)\nonumber\\
& \Leftrightarrow & r_1^2=2\ln\left(\frac{a^4N}{(2\pi)^2}\right)-r_2^2=B-r_2^2
\label{cond}
\end{eqnarray}
with $B=2\ln\left(\frac{a^4N}{(2\pi)^2}\right)$. Hence a necessary condition for $P^{(1)}(\mathbf{x_1}, \mathbf{x_2})>1$ is that both $r_1<\sqrt{B}$ and $r_2<\sqrt{B}$. The degree distribution is then obtained from Eq. \ref{distr} (see Appendix for the complete derivation):
\begin{eqnarray}
P(k)\sim \left\{\begin{array}{ccc}
\text{const} & \text{if} & r<\sqrt{B} \Leftrightarrow k>\frac{2\pi}{a^2}\\
\frac{1}{k} & \text{if} & r>\sqrt{B} \Leftrightarrow k<\frac{2\pi}{a^2}
\end{array}
 \right.
\label{deg2}
\end{eqnarray}
Note that the flat tail for $k>2 \pi /a^2$ is an artifact of the continuous approximation in $D=2$. Analytical calculations, for instance in $D=4$ where they are still manageable (but somehow tedious), show a decreasing behavior even for $k>2\pi/a^2$.

In the same way,  the average neighbor connectivity is computed from Eq. \ref{knn.k} and the clustering coefficient from Eq. \ref{c.k}. Results are shown graphically on Fig. \ref{fcor} (see Appendix for analytical calculations).

In order to compare the analytical predictions  obtained above for $n=1$ with the CSN obtained from simulations, a Langevin dynamics with potential energy defined by Eq. \ref{uxy} is performed according to the equation of motion:
\[\gamma \dot{\mathbf{x}}=-\frac{\partial U}{\partial \mathbf{x}}+f(t)\]
where $\gamma$ is the friction coefficient and $f(t)$ is a white noise with mean value $<f(t)>=0$ and $<f(t)f(t')>=\delta(t-t')$.
Without loss of generality $\gamma$ is set to one (it merely corresponds to a rescaling of the time scale) and the integration step to $i_t=0.001$ in all simulations.
In the case of the two-dimensional quadratic well, configurations are defined as square cells of size $a^2$ ($a$=0.2). A total number of $N=3\cdot 10^6$ time steps has been employed for all simulations.

The degree distribution $P(k)$ for different values of the parameter $M$ is shown in Fig. \ref{fcor}A. The distribution follows a power-law of the form $1/k$ for values of the parameter $M\geq 100$. 
In this example, all the CSN realizations with $M\geq 10^{4}$ are equivalent to a random sampling of the energy landscape with probabilities given by $W_oe^{-U(x)}$ (black points in the figure). Hence for these values of $M$ the distribution follows the analytical prediction.

In Figure \ref{fcor}B, the average neighbor connectivity $K_{nn}(k)$ is plotted for different values of $M$. There is a change in the behavior of $K_{nn}(k)$ as $M$ increases. For low values of $M$ $K_{nn}(k)$ is an increasing function of $k$, which indicates an assortative behavior. This is no longer true for large values of $M$. In this regime, $K_{nn}(k)$ shows a decaying tail typical of disassortative regime. For $M\ge 10^4$ the curves cannot be distinguished from the one obtained with uncorrelated sampling. The flat region for small $k$ arises because nodes with small degree are likely to be connected with nodes with high degree which lay at the bottom of the basin. This phenomenon reflects that for large $M$, transitions starting at a node far from the minimum are likely to end up at the bottom of the basin, which is characterized by nodes with a large degree. On the other hand, for small values of $M$ only neighbor configurations are visited consecutively. 

\begin{figure}[htbp]
  \begin{center}
\includegraphics[width=8.0cm]{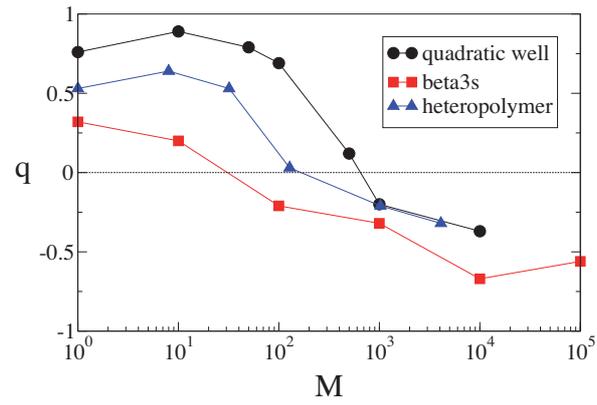}
\caption{(Color online) Assortativity mixing coefficient for different values of the parameter $M$ for the three systems under study.}
\label{ass.fig}
\end{center}
\end{figure}

As already pointed out, the approximation $n=1$ has the effect of slightly overestimating the node degree, which explains why results for the uncorrelated sampling are slightly below the analytical approximation.
The calculation of the network assortativity coefficient $q$ \cite{Newman2002} for different values of $M$ reveals the same transition between  assortative and  disassortative regimes (see Fig. \ref{ass.fig}). For $M<100$ the network presents a strong assortativity characterized by values around $q\approx 0.8$. Increasing the value of $M$ makes the assortativity coefficient drop to values smaller than $-0.3$ indicating that the system has undergone an assortative-to-disassortative transition.

Although models with changing assortativity have been recently proposed \cite{wang2006} this is the first time to our knowledge that a family of networks underlying the same physical process,  i.e.  diffusion in a  well, presents this kind of transition.

In the same way, the clustering coefficient $C(k)$ exhibits different behaviors as a function of $M$. For $M<1000$ the value of $C(k)$ grows indicating that triangles easily form at the bottom of the basin. On the other hand, as $M$ increases, $C(k)$ shows a decaying tail for large values of the degree $k$. For $M>10^{4}$ the $C(k)$ obtained from the Langevin dynamics follows the analytical prediction of Eq. \ref{clust}. 

\begin{figure}[htbp]
\begin{center}
\includegraphics[width=160mm]{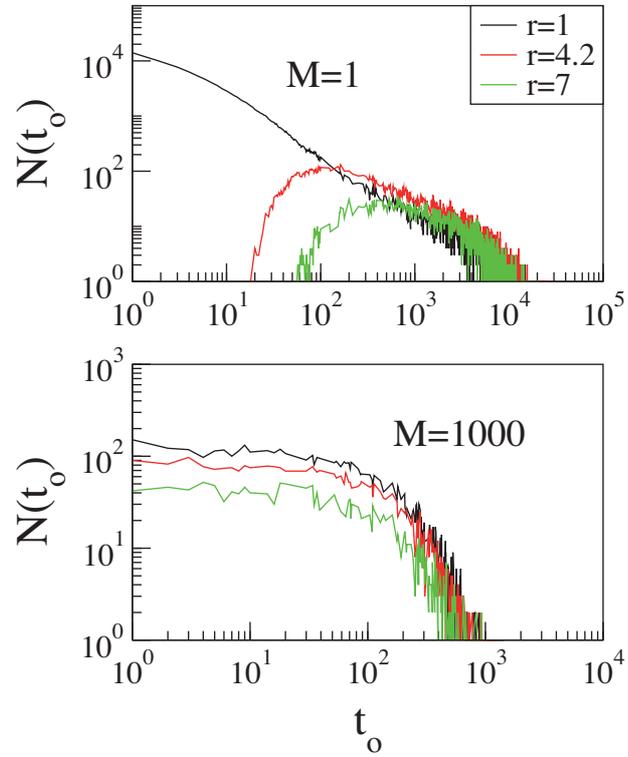}
\caption{(Color online) Distribution of the relaxation times from three different initial configurations to the bottom configuration of the quadratic well in $D=2$ dimensions (A) for $M=1$ and (B) for $M=1000$. The value $r$ indicate the radial distance from the starting node for each of the three curves.}
\label{trelax.fig}
\end{center}
\end{figure}

The changing behavior of the CSN topology of a quadratic well for different values of the configuration saving time can be understood in a more general kinetic framework when considering relaxation times to a given configuration of the landscape. In Fig. \ref{trelax.fig}A the distribution of the relaxation times to the configuration laying at the bottom of the well for three different configurations is shown. Configurations at a small distance from the bottom node (small $r$) exhibit a downhill distribution, which is not the case for larger values of $r$. However as $M$ increases, all distributions overlap (up to a global multiplicative factor) indicating that the kinetics to the bottom is the same  for all configurations (see Fig. \ref{trelax.fig}B). This corresponds to the uncorrelated regime of Eq. \ref{wij}. In this case, the probability to have a link between two configurations depends only on the configuration weight.

\section{Native state of a triple stranded $\beta$-sheet and lattice heteropolymer}

\begin{figure}[htbp]
\begin{center}
\includegraphics[width=16.0cm]{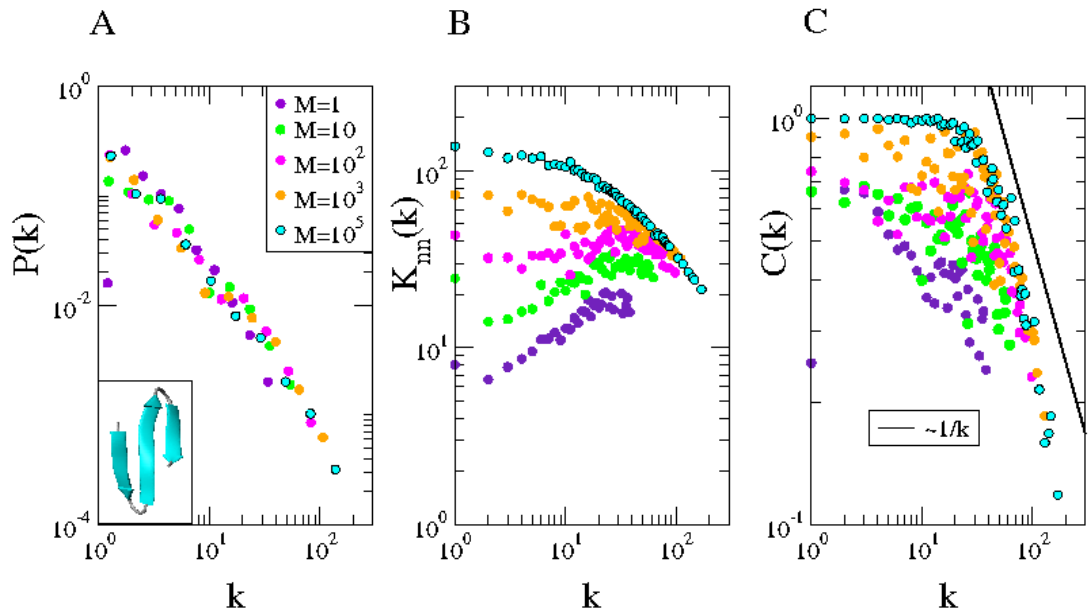}
\caption{(Color online) Network topology for the beta3s peptide CSN at different values of the parameter $M$. (A) Degree distribution. To reduce noise a logarithmic binning has been applied. The native state of beta3s is shown in the inset. (B) Average neighbor connectivity. (C) Clustering coefficient.}
\label{topo.fig}
\end{center}
\end{figure}

\begin{figure}[htbp]
\begin{center}
\includegraphics[width=16.0cm]{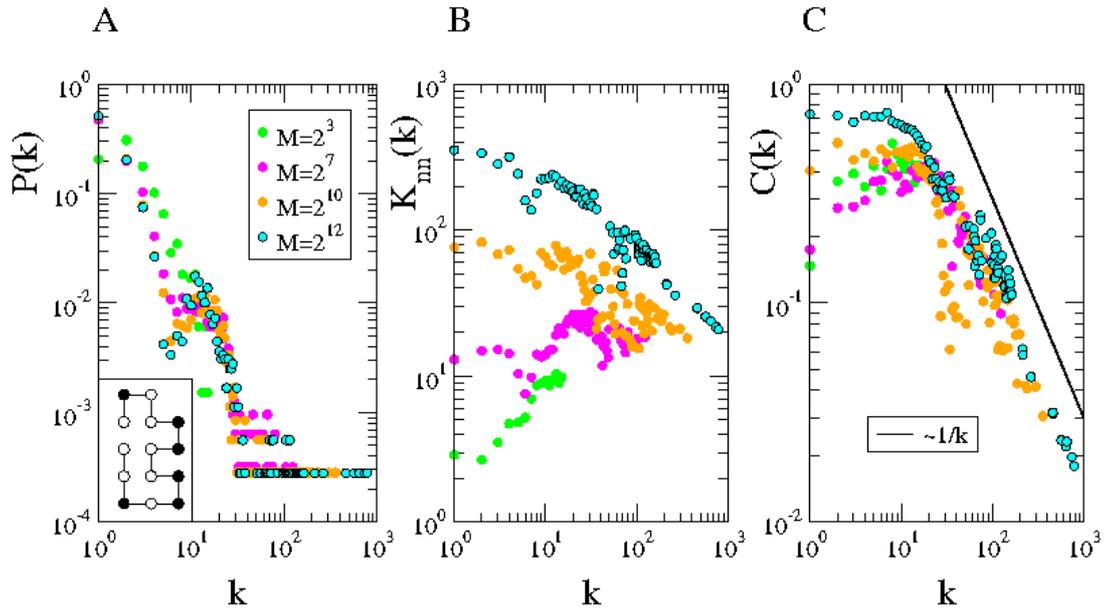}
\caption{(Color online) Network topology for the random lattice heteropolymer CSN at different values of the parameter $M$. (A) Degree distribution. The most visited configuration of the heteropolymer is shown in the inset. (B) Average neighbor connectivity. (C) Clustering coefficient.}
\label{Mfcor}
\end{center}
\end{figure}

The analytical and numerical results obtained above are crucial for a correct interpretation of the CSN topology observed in complex systems for which direct application of Eq. \ref{distr}-\ref{c.k} is unfeasible. In the following, the CSN topology of the native basin of a triple stranded $\beta$-sheet peptide (beta3s) sampled by MD as well as of a
lattice heteropolymer sampled by MC are investigated (see Fig. \ref{topo.fig} and \ref{Mfcor}). 

The MD simulation of the native state  of beta3s is performed at 270 K for a total of 10 ns of simulation time which is enough for the correct sampling of the basin. The low simulation temperature prevented the system from jumping to a different basin. The MD simulation is performed using the CHARMM PARAM19 force field \cite{Brooks1983} and an integration time step of $i_t=2\ f$s. A mean field approximation based on the solvent-accessible surface was used to describe the main effects of the aqueous solvent on the solute \cite{Ferrara2002}. The two parameters of the solvation model were optimized without using beta3s. The same force field and implicit solvent model have been used recently in MD simulations of various systems \cite{Ferrara2000,Cavalli2002,Gsponer2003}. 

Secondary structure is calculated \cite{Carter2003} for each snapshot saved along the MD trajectory. Here a configuration (i.e., a CSN node) is defined as a single string of secondary structure e.g., the most populated configuration for beta3s at 270 K (see inset of Fig. \ref{topo.fig}) is: -EEE-STTEEEEESSEEEE-. The total number of $5\times 10^6$ snapshots sampled during the MD simulation resulted in 249 secondary structure configurations. There are eight possible letters in the secondary structure ``alphabet'': ``H'', ``G'', ``I'', ``E'', ``B'', ``T'', ``S'' and ``-'' standing for $\alpha$-helix, $3_{10}$ helix, $\pi$-helix, extended, isolated $\beta$-bridge, hydrogen bonded turn, bend, and unstructured, respectively. Since the N and C-terminal residues are always assigned an ``-'' a 20 residue peptide can, in principle, assume $8^{18}\approx10^{16}$ configurations. 

The 2D lattice heteropolymer is simulated in the framework of the popular HP
model \cite{Dill1985,Sali1994,Dill1995,Karplus1995,Thirumalai1996}. In this description, the amino acid
sequence of a protein is represented as a binary sequence of hydrophobic (H)
and polar (P) residues. The results presented here are obtained with
the random sequence HHPHPPHHPPHHPH (inset of Fig. \ref{Mfcor}). Note that similar results are observed for different HP
sequences (random and protein-like) as well as for different numbers of beads, ranging from $10$
to $20$ (a detailed presentation of these results is under preparation). The
time series of configurations is generated from moves of the polymer
according to the Metropolis rule. It is important to mention that the
qualitative observations do not depend on the set of moves (local moves
and the global "pivot" moves \cite{Sokal1995, Madras1996} have been tested). In the standard HP model, the energy of a
configuration is merely minus the number of its H-H contacts on the lattice.
From a physical point of view, the cornerstone of the simulation is the
appropriate adjustment of $T$ in order to achieve an effective sampling of
the lowest energy configurations. This has been accomplished by sampling at a temperature
significantly smaller than the coil-to-globule transition temperature of the
polymer (i.e. $k_BT_{samp}=0.3$ and $k_BT_{trans}\approx 0.5$). The transition
temperature has been identified by a thorough study of the heat capacity $%
C_{V}$ and of two topological quantities, namely the gyration radius and the
end-to-end distance. A CSN node is defined as a single lattice configuration up to a symmetry of the lattice.

In both the beta3s and heteropolymer systems two nodes are linked if a direct transition between them (at a given $M$) has been observed during the simulation. The topology of the two CSNs shows several common properties. The degree distributions $P(k)$  for beta3s and the heteropolymer are shown in Fig.\ref{topo.fig}A and \ref{Mfcor}A, respectively. The distributions are robust upon varying the configuration saving time (i.e. changing the value of $M$) and resemble a power-law $k^{-\gamma}$ for $M\ge 1$ with exponent $\gamma$ between 1.5 and 2. This behavior is qualitatively similar to what is observed for the quadratic well while the steeper slope may result from the higher dimension (i.e., higher number of degrees of freedom) of the energy landscapes. Interestingly the average neighbor connectivity $K_{nn}(k)$ changes significantly for different values of $M$. $K_{nn}(k)$ is shown in Fig. \ref{topo.fig}B and \ref{Mfcor}B. For $M<100$ this quantity grows with the degree, whereas for larger values of $M$, $K_{nn}(k)$ becomes a decreasing function of $k$.
Moreover, the assortativity coefficient $q$ changes from positive values for $M=1$ to negative values for $M>1000$ indicating an assortative-to-disassortative transition (see Fig. \ref{ass.fig}). 

The clustering coefficient $C(k)$ converges towards a general decreasing behavior for large $M$ (Fig. \ref{topo.fig}C and \ref{Mfcor}C).
In a previous work, the presence of an apparent scaling in $C(k)$ had been interpreted as the signature of a hierarchical organization of the nodes in the native state of beta3s \cite{Rao2004}. However, a comparison between the $C(k)$ of beta3s for different values of $M$ and of the quadratic well (see Fig. \ref{fcor}) strongly suggest that this decay does not indicate node hierarchy as presented in \cite{Ravasz2003}. Firstly, the quadratic well underlying the CSN does not present a hierarchical organization like in \cite{Ravasz2003}. Secondly, it should be noticed that in the uncorrelated regime nodes laying at the bottom of the basin are connected together, giving rise to an almost complete subgraph. In this regime, nodes with low degree are unlikely to be connected together but tend to connect to high degree nodes (bottom configurations). These two effects are indeed sufficient to explain the decay observed in $C(k)$ without invoking the presence of a node hierarchy.

\begin{figure}[htbp]
\begin{center}
\includegraphics[width=8.0cm]{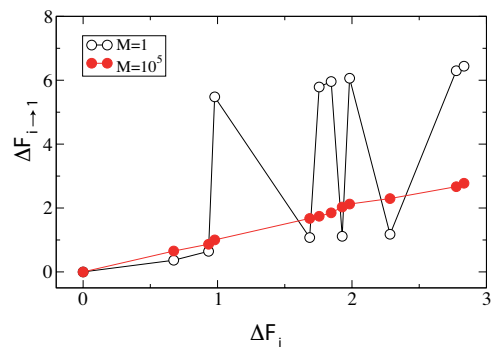}
\caption{(Color online) Relation between the free-energy barrier to the configuration at the bottom of the native state basin ($\Delta F_{i\rightarrow 1}$), and the configuration free energy ($\Delta F_i$) for the most visited nodes of the beta3s network. Empty and full dots represent the $M=1$ and $M=10^5$ case, respectively.}
\label{wmtx.fig}
\end{center}
\end{figure}

For the CSN of beta3s there exists no rigorous evidence that for $M>100$ an uncorrelated regime is reached. However, analysis of the transition probabilities (i.e. link weights) can account for this behavior. In Fig. \ref{wmtx.fig} the relation between $\log (w_i/w_{1})$ and $\log(w_{i\rightarrow 1}/w_{2\rightarrow 1})$ is shown, where $w_i$ and $w_{i\rightarrow 1}$ indicate the weight of node $i$ and the weight of the link between node ``$i$" and ``1", respectively. Index 1 stands for the most populated node of the network. These logarithms have a physical meaning reflecting the configuration free-energy 
$\Delta F_{i}\approx -k_{B}T\log(w_{i})$ 
and the free-energy barrier between different configurations 
$\Delta F_{i\rightarrow j}\approx -k_{B}T \log(w_{i\rightarrow j})$. 
For $M=1$ the relation between the two free-energies is not linear. In other words, nodes with similar weights might be separated from node ``1'' by free-energy barriers of very different size (for instance the two nodes with $\Delta F_{i}\approx 1$ in Fig. \ref{wmtx.fig}). Choosing higher $M$ increases the correlation between node and link weights. For $M=10^5$, $\Delta F_{i\rightarrow j}$ grows linearly with  $\Delta F_{i}$ indicating that link weights depend only on $w_{i}$. This behavior provides strong evidence for an uncorrelated sampling.

It is essential to stress that the uncorrelated node regime is a frequent scenario when dealing with long sampling MD studies. These simulations explore transitions between several energy basins, for example, when investigating the large configurational changes characterizing protein folding. In these cases, the configuration saving time is usually set to large values for computational reasons resulting in an intra-basin uncorrelated regime.
Finally, it is important to note that these results have been obtained for CSNs originating from a single basin energy landscape. In the case of networks describing fully sampled landscapes presenting a large number of basins the network topology reflects the contributions from different basins.

\section{Connection with fitness (hidden variable) models}
\label{fitness}

The scaling behavior observed in several networks has triggered a vast effort in modeling complex networks \cite{Barabasi1999}. Of particular interest for CSN is the model based on a fitness parameter on the nodes  \cite{Caldarelli2002, Boguna2003, Servedio2004}.
In the original fitness model \cite{Caldarelli2002} two nodes are connected with probability one if the sum of their fitness exceeds a given threshold. In the case of CSNs reflecting a single enthalpic energy basin (like in this work), the fitness of a node is given by $-U(\mathbf{x})$. Eq. \ref{unif} with $n=1$ shows that nodes are connected with probability one if the sum of their fitness is higher than a threshold given by $-\ln(W_0^2a^{2D}N)$. In addition there is also a remaining probability to connect nodes of high energy, given by $NP(\mathbf{x_2})P(\mathbf{x_1})$.
This formulation shows that CSNs fall in the large class of networks whose nodes are described by a fitness parameter, also referred to as hidden variable \cite{Boguna2003}. Notably, the behavior of $P(k)$, $K_{nn}(k)$ and $C(k)$ in the uncorrelated case is in good agreement with those obtained in the previous works mentioned above.

\section{Conclusions}

The scaling behavior observed in the CSN topology has been investigated in the quadratic-well model, the native state of a triple stranded $\beta$-sheet peptide and a lattice heteropolymer model. Three main results have clearly emerged.
Firstly, in the limit of very large configuration saving times (uncorrelated regime) an analytical approximation ($1^{\text{st}}$ order) for the degree distribution, the average neighbor connectivity and the clustering coefficient is provided. Comparison between the analytical predictions and the results obtained from the simulation of the dynamics in a quadratic well shows that, in the limit considered, the analytical solution describes correctly the CSN topology. These results allow for the interpretation of the topology observed in complex CSNs which cannot be tackled analytically, like the ones describing the native state of a $\beta$-sheet peptide or the low-energy configurations of a lattice heteropolymer.
Secondly, the variation of the configuration saving time induces remarkable changes in the CSN topology. For small saving times the network exhibits an assortative mixing. On the other hand, when increasing the saving time a disassortative behavior is observed. 
Given the same physical process (i.e. diffusion in a well), this result shows how the associated CSN changes its topological properties.
Thirdly, the emergence of a decaying tail in the clustering coefficient, which had been suggested to bear the signature of a hierarchical organization of the nodes in the native state of the $\beta$-sheet peptide, is in fact a consequence of uncorrelated sampling.

\section*{Acknowledgments}

F.R. thanks A. Caflisch, M. Karplus and S. Krivov for a critical reading of the manuscript. D.MDL. thanks Maciej Kurant for fruitful discussions. D.G. acknowledges the financially support of COSIN (FET Open IST 2001-33555), DELIS (FET Open 001907) and the SER-Bern (02.0234).

\section*{Appendix}

For the case of the quadratic well in $D=2$, the derivation of the degree distribution (Eq. \ref{deg2}) is performed by first calculating the degree of a node at distance $r$.
If $r\leq \sqrt{B}$, Eq. \ref{deg} reads:
\begin{eqnarray}
k(r)&=&\frac{2\pi}{a^2}\int_0^{\sqrt{B-r^2}}r_1\text{d}r_1+\frac{2\pi}{a^2}\int_{\sqrt{B-r^2}}^{\infty}r_1\text{d}r_1\frac{a^4N}{(2\pi)^2}e^{-\frac{1}{2}(r_1^2+r^2)}\nonumber\\
&=&\frac{2\pi}{2a^2}(B-r^2)+\frac{a^2N}{2\pi}e^{-\frac{1}{2}B}\nonumber\\
&=&\frac{2\pi}{2a^2}(2+B-r^2)
\label{r1}
\end{eqnarray}
If $r>\sqrt{B}$, Eq. \ref{deg} reads:
\begin{eqnarray}
k(r)&=&\frac{2\pi}{a^2}\int_0^{\infty}r_1\text{d}r_1\frac{a^4N}{(2\pi)^2}e^{-\frac{1}{2}(r_1^2+r^2)}\nonumber\\
&=&\frac{a^2N}{2\pi}e^{-\frac{1}{2}r^2}
\label{r2}
\end{eqnarray}
Eq. \ref{distr} is calculated using the properties of the $\delta(f(r))$ function. For a given function $f(r)$ with $n$ simple zeros $f(r_{i}^{*})=0,f^{'}(r_{i}^{*})\neq 0, i=1\dots n$, it is possible to write:
$\delta(f(r))=\sum_{i=1}^{n} \frac{\delta(r-r^{*}_{i})}{|f^{'}(r^{*}_{i})|}$
Hence $r*$ is given by inverting Eq. \ref{r1} and \ref{r2}:
\begin{enumerate}
\item if $r<\sqrt{B} \Leftrightarrow k>\frac{2\pi}{a^2}$: 
\[r^*=\sqrt{2(1+\frac{B}{2})-\frac{a^2}{\pi}k}\]
\[\Leftrightarrow P(k)\sim \frac{r^*}{\left|\frac{2\pi}{a^2}r^*\right|}\sim \text{const}\]
\item if $r<\sqrt{B} \Leftrightarrow k<\frac{2\pi}{a^2}$: 
\[r^*=\sqrt{2\ln \frac{a^2N}{2\pi k}}\]
\[ \Leftrightarrow P(k)\sim \frac{r^*}{\frac{a^2N}{2\pi}r^*e^{-\frac{1}{2}(r^*)^2}}\sim\frac{1}{k}\]
\end{enumerate}
The average neighbor connectivity is calculated using Eq. \ref{knn.k}:

\begin{eqnarray}
K_{nn}(k)= \left\{\begin{array}{ccc}
\frac{2\pi^2}{a^4}2\frac{1}{k}+\frac{2\pi}{a^2}(1+\frac{B^2}{2})-\frac{1}{2}k+\frac{\pi}{a^2}e^{-1-\frac{B}{2}}\frac{1}{k}e^{\frac{a^2}{2\pi k}} & \text{if} & k>\frac{2\pi}{a^2}\\
B\frac{\pi}{a^2}+\frac{4\pi^3}{a^6N} & \text{if} & k\leq \frac{2\pi}{a^2}
\end{array}
 \right.
\label{knn2}
\end{eqnarray}
The expression for the clustering coefficient is slightly more complex since it requires to distinguish between several cases according to the possible values of $r$. Of particular interest is the case of large $k$ (i.e. small $r$). If $r<\sqrt{\frac{B}{2}}\Leftrightarrow k>\frac{2\pi}{a^2}(1+\frac{B}{4})$, solving the integral of Eq. \ref{tr} gives:
\begin{eqnarray}
C(k)=\frac{N^{\bigtriangleup}(k)}{k^2}&=&\frac{(2\pi)^2}{a^4}\frac{1}{k^2}\left[\frac{1}{2}B+\frac{B^2}{8}-1-\left(\frac{a^2}{2\pi}k-1-\frac{B}{2}\right)^2\right]\nonumber\\
&+&\frac{(2\pi)^2}{a^4}\frac{1}{k^2}\left[\exp\left(\frac{a^2}{2\pi}k-1-\frac{B}{2}\right)+\frac{\pi^2}{a^4}\exp\left(-\frac{a^2}{\pi}k+2+B\right)\right]
\label{clust}
\end{eqnarray}

\bibliographystyle{apsrev}

\clearpage

\begin{figure}[htb]
  \begin{center}
    \includegraphics[width=160mm]{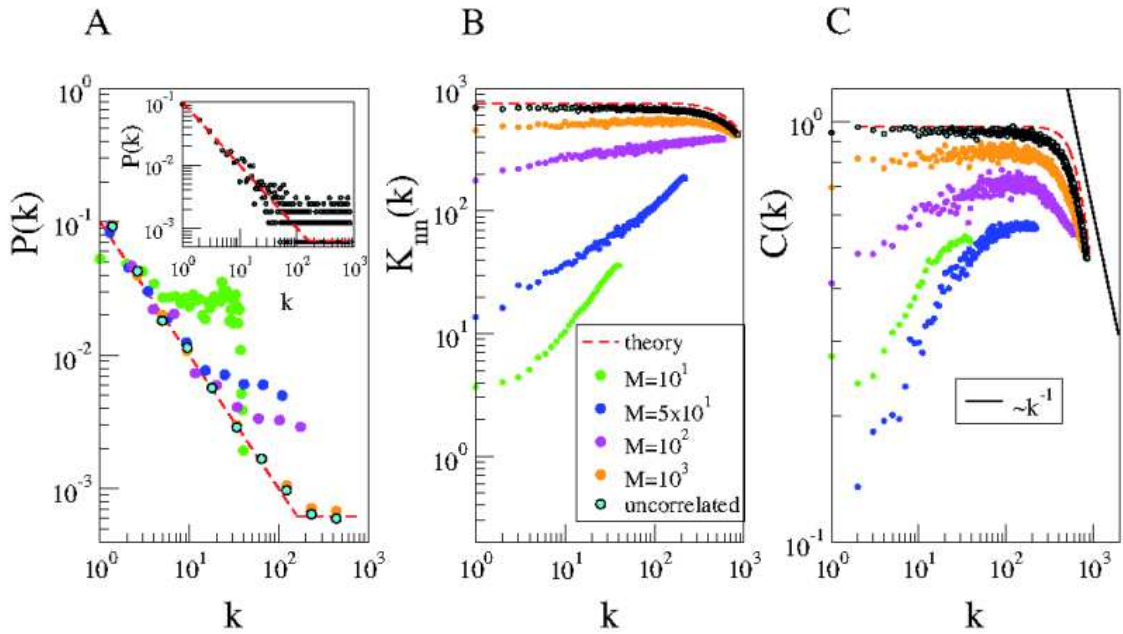}
\caption{(Color online) Network topology for the quadratic well in $D=2$ dimensions and different values of the parameter $M$. (A)~Degree distribution. For clarity a binning has been applied for $M>10$. Inset: degree distribution for uncorrelated sampling without binning. (B)~Average neighbor degree. (C)~Clustering coefficient. Black dots are obtained by a random sampling of the energy landscape (uncorrelated case, see text). Red dashed-lines shows the analytical estimation.}
    \label{fcor}
  \end{center}
\end{figure}

\end{document}